\documentclass[aps,prl,twocolumn,eqsecnum,amssymb,amsmath,showpacs,a4paper, superscriptaddress]{revtex4-2}

\usepackage{graphicx}
\usepackage{amsfonts}
\usepackage{amsmath}
\usepackage{hyperref}
\usepackage{units}
\usepackage{color}
\usepackage{dcolumn}
\usepackage{bm}
\usepackage{float}
\usepackage{cleveref}
\usepackage[normalem]{ulem}
\usepackage{appendix}
\usepackage{mathtools}
\usepackage{esvect}

\usepackage[colorinlistoftodos, size=tiny, bordercolor=white]{todonotes}
\usepackage{comment}

\usepackage[subpreambles=true]{standalone}  
\usepackage{import} 

\graphicspath{ {images/} }
\usepackage{mathrsfs}
\usepackage{amssymb}
\usepackage{dsfont}
\usepackage{enumitem}
\usepackage{gensymb}
\usepackage{bm}

\crefname{section}{§}{§§}
\Crefname{section}{§}{§§}
\usepackage{mathtools}
\usepackage{chngcntr}
\counterwithout{equation}{section}

\usepackage{tikz} 
\usetikzlibrary{shapes,arrows,positioning,automata,backgrounds,calc,er,patterns}
\usepackage[compat=1.1.0]{tikz-feynman}

\makeatletter
\let\cat@comma@active\@empty
\makeatother
\newcommand{\tp}{\mathit{tp}}

\newcommand{\sr}{\mathit{sr}}

\newcommand{\cmmnt}[1]{\ignorespaces}

\newcommand{\be} {\begin{equation}}
\newcommand{\ee} {\end{equation}}
\newcommand{\bsub}{\begin{subequations}}
\newcommand{\esub}{\end{subequations}}
\newcommand{\bea}{\begin{eqnarray}}
\newcommand{\eea}{\end{eqnarray}}

\begin{document}

\title{The sound-ring radiation of expanding vortex clusters}
\author{August Geelmuyden}
\affiliation{School of Mathematical Sciences, University of Nottingham, University Park, Nottingham, NG7 2RD, UK}

\author{Sebastian Erne}
\affiliation{School of Mathematical Sciences, University of Nottingham, University Park, Nottingham, NG7 2RD, UK}
\affiliation{Vienna Center for Quantum Science and Technology, Atominstitut, TU Wien, Stadionallee 2, 1020 Vienna, Austria}

\author{Sam Patrick}
\affiliation{School of Mathematical Sciences, University of Nottingham, University Park, Nottingham, NG7 2RD, UK}
\affiliation{Department of Physics and Astronomy, University of British Columbia, Vancouver, British Columbia, V6T 1Z1, Canada}

\author{Carlo Barenghi}
\affiliation{Joint Quantum Centre Durham-Newcastle, School of Mathematics, Statistics and Physics, Newcastle University,
Newcastle upon Tyne, NE1 7RU, UK}

\author{Silke Weinfurtner}
\affiliation{School of Mathematical Sciences, University of Nottingham, University Park, Nottingham, NG7 2RD, UK}
\affiliation{Centre for the Mathematics and Theoretical Physics of Quantum Non-Equilibrium Systems, University of Nottingham, Nottingham, NG7 2RD, UK}

\date{\today}

\begin{abstract}
\noindent
We investigate wave-vortex interaction emerging from an expanding compact vortex cluster in a two-dimensional Bose-Einstein condensate. 
We adapt techniques developed for compact gravitational objects to derive the characteristic modes of the wave-vortex interaction perturbatively around an effective vortex flow field. 
We demonstrate the existence of orbits or sound-rings, in analogy to gravitational light-rings, and compute the characteristic spectrum for the out-of-equilibrium vortex cluster.
The spectrum obtained from numerical simulations of a stochastic Gross-Pitaevskii equation exhibiting an expanding vortex cluster is in excellent agreement with analytical predictions.
Our findings are relevant for 2d-quantum turbulence, the semi-classical limit around fluid flows, and rotating compact objects exhibiting discrete circulation. 
\end{abstract}

\maketitle
{\textbf{Introduction.}}---
The interaction between waves and rapidly rotating objects gives rise to a range of exotic phenomena. A prominent example is that of compact gravitational objects, such as black holes and neutron stars, which can exhibit circular orbits of light commonly referred to as light-rings~\cite{cardoso2009geodesic}. The existence of light-rings implies that the gravitational field significantly affects the wave propagation nearby. In particular, the light-ring is intimately connected to the quasi-normal modes of black holes and neutron stars~\cite{berti2009review}, which are the damped linear oscillations (ringdown) of a system as it settles into equilibrium. This black hole relaxation process, or ringdown, following a black hole merger event has been observed by the LIGO collaboration~\cite{abbott2016merger,abbott2017observation}. 

The importance of the light-ring in characterising scattering processes is not limited to gravitational systems. One can identify analogue or effective light-rings in hydrodynamic systems. To derive effective light-rings in the system one studies the wave-propagation of bulk (i.e.~sound waves) or interface waves~\cite{torres2019analogue} in rapidly rotating fluids. Recently it was shown that the free surface of a macroscopic classical vortex flow oscillates at the analogue of light-ring frequencies~\cite{torres2020quasinormal}. 

A promising new line of research within the gravitational wave community is to extend the concept of light-rings from beyond the aftermath of a binary merger all the way back before the merger begins~\cite{mcwilliams2019analytical}. This is somewhat surprising since the light-ring is tied to wave scattering around a stationary gravitational background, whereas a binary merger refers to a highly non-linear gravitational field evolution. If the light-ring has predictive power for the non-linear dynamics in general relativity, might the same then be true in hydrodynamic systems? We show that this is the case, by adapting the light-ring methods to predict the spectrum of sound emitted by a rapidly decaying compact vortex cluster in a two-dimensional Bose-Einstein condensate (BEC). Here, the effective light-rings correspond to circular trajectories of sound, or \textit{sound-rings}. 

\begin{figure} 
\centering
\includegraphics[width=\linewidth]{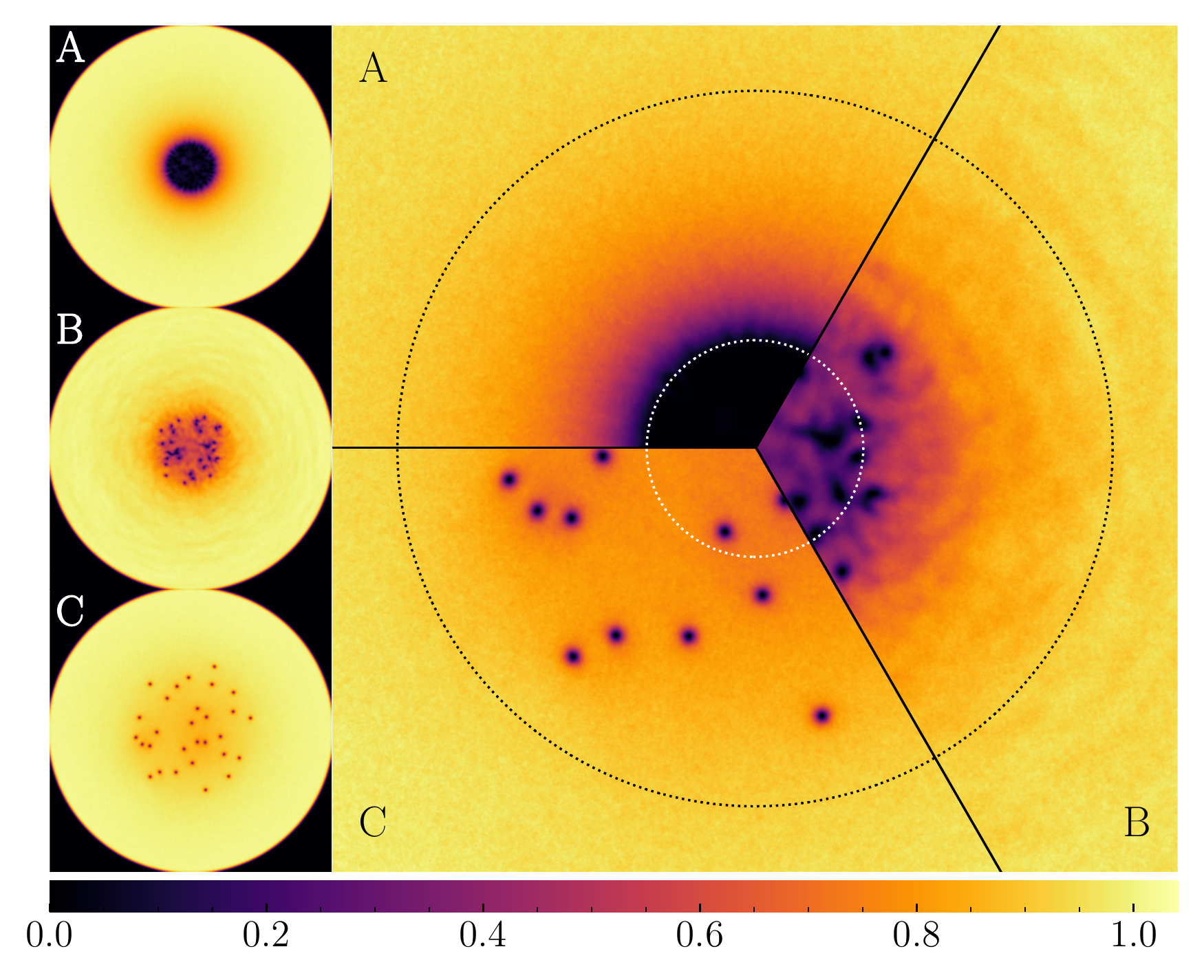}
\caption{\textbf{Simulations of a decaying vortex cluster.} Left column: condensate density at three qualitatively different stages of the decay. initial multi-winded vortex (A); multi-winded vortex has decayed in many singly winded vortices forming a disordered cluster (B); cluster has expanded (C). Right column: enlargements to show the vortex configuration at the three stages together with the outermost sound-ring radius (black circle) and the minimal cluster size (white circle). }
\label{fig:sim}
\end{figure}

Being a macroscopic quantum fluid, a BEC is known to exhibit a wide range of exotic phenomena including wave-propagation of heat (second sound)~\cite{andrews1997propagation}, perfectly inviscid flow, and quantization of circulation~\cite{barenghi2016primer}. The latter leads to vortices in BECs (holes in the condensate) that are discretized and topologically protected by their phase winding. 
Whereas prior studies of effective light-rings have been conducted in draining (spiral) classical hydrodynamical vortices~\cite{torres2020quasinormal}, we show that the notion of light-ring extends to the purely circular flow around quantized vortices and clusters thereof. As we will show, the sound-ring of a cluster can extend well outside the cluster core. Consequently, the cluster exhibits a characteristic spectrum independent of vortex configuration. This demonstrates the predictive power of sound-rings for composite rotational systems. 
Our findings are of direct relevance for wave-vortex interaction in superfluids, complementing ongoing research in quantum turbulence in two and three dimensions exhibiting clusters and bundles of vortices, respectively.

{\textbf{Modelling.}}---
We investigate the non-linear relaxation of a rapidly rotating disc-shaped BEC through numerical simulations of the two-dimensional (2D) Stochastic Gross-Pitaevskii Equation (SGPE),
\cite{Cockburn_2009,Proukakis_2008}, 
\begin{equation} \label{SGPE}
\begin{split}
    i\hbar \partial_t \Psi = & \ \left(1-i\gamma\right)\left[-\frac{\hbar^2}{2m}\nabla^2+U(\mathbf{r})-\mu+g|\Psi|^2\right]\Psi \\ 
    & \quad + \eta(\mathbf{r},t) ~.
\end{split}
\end{equation}
Here, $t$ is time and $\mathbf{r}=(r,\theta)$ are 2D polar coordinates. We consider a tight transverse confinement, such that dynamics along the the transverse $z$ direction is effectively frozen. The wavefunction $\Psi$ describes the low-energy coherent modes in the system, coupled to a thermal bath through the complex noise $\eta$ and damping $\gamma$~\cite{SM}. Here $m$ is the boson mass, $\mu$ is the chemical potential, $g$ is the effective 2D interaction constant, and $U(\mathbf{r})$ is a rotationally symmetric box-trap with $U(\mathbf{r})\gg \mu$ for $|\mathbf{r}| \geq r_B$, where $r_B$ is the radius of the confining potential~\cite{SM}. In the following, we consider dimensionless variables, with length measured in units of the healing length $\xi=\hbar/\sqrt{m\mu}$, time in units of $\mu/\hbar$, and the wavefunction in units of $\sqrt{\mu/g}$. In these units, only the external potential $U$ and noise $\eta$ depend on the microscopic details of the system such that solutions to Eq.~\eqref{SGPE} are applicable over a wide range of parameters.

Fig.~\ref{fig:sim} illustrates a typical evolution for the relaxation of a vortex cluster. We initialize a cluster of singly winded vortices through the decay of an energetically and dynamically unstable multiply ($\ell=29$) winded vortex (stage A) \cite{giacomelli2020ergoregion}. Here, the stochastic driving $\eta$ seeds the instability, ensuring spatial disorder of the resulting vortex ensemble by explicitly breaking rotational symmetry. The initial nonlinear relaxation of the cluster shows a strong increase in fluctuations transferring energy away from the core (stage B). Here, the individual vortices form common time-independent sound-rings extending to a radius $r_\sr$ far outside the cluster. These sound-rings determine the low-frequency sound-field outside the cluster, independent of the nonlinear dynamics within its core. During the subsequent evolution (stage C) the cluster further expands under the influence of the damping $\gamma$, which (i) reduces reflection of fluctuations from the boundaries and (ii) enables us to tune the typical timescale of the expansion. Note, the presented results are independent of their specific values for sufficiently small $\eta$ and $\gamma$~\cite{SM}.
Once vortices have expanded beyond the largest $r_\sr$ the common sound-ring disappears, breaking up into individual sound-rings for each vortex. This removes the potential barrier enabling the low-frequency sound-field to probe the cluster core.

\textbf{Vortex cluster.}--- We study the fluctuations in the hydrodynamic approximation, writing the wavefunction as $\Psi = \sqrt{\rho}e^{i\Phi}$, and consider small fluctuations of the density $\rho = \rho_0 + \delta \rho$ and phase $\Phi = \phi_0 + \delta \phi$, around a coarse-grained background field $\psi_0 = \sqrt{\rho_0} e^{i\phi_0}$. The background $\psi_0$ includes the macroscopic flow of the condensate and, in general, depends on the motion of individual vortices. However, for a sufficiently high number of vortices we can define a slowly evolving coarse-grained background, by time-averaging over the fast rotation of the cluster around its own axis. Therefore, $\psi_0$ only evolves due to the slow expansion of the cluster.

Chaotic movement of vortices within the cluster ensures ergodicity, such that we can determine the time-averaged background through ensemble averages~\cite{SM}.
For a collection of $N$ irrotational vortices with winding $\ell_j$, uniformly distributed within a cluster of size $R$, the coarse-grained velocity field $\mathbf{v}\equiv \nabla \phi_0=v_\theta(r)\,\hat{\mathbf{e}}_\theta$ is given by the Rankine vortex
\begin{equation} \label{vel}
v_\theta(r) = \begin{cases}
\ell r/R^2, \quad r\leq R \\
\ell/r, \qquad \  r>R
\end{cases}
\text{ for } 
\ell \equiv \sum_j^N \ell_j ~,
\end{equation}
familiar from classical fluid mechanics \cite{lautrup2011physics}. In accordance with Feynman's rule \cite{barenghi2016primer}, the flow within the cluster ($r<R$) resembles solid body rotation. Far outside the cluster ($r>R$) one recovers, even without the time-averaging, the irrotational flow field of a single multiply winded vortex. Note, in particular, that this is independent of the details within the core and only depends on the net winding $\ell$ of the cluster~\cite{SM}. 

In the Thomas-Fermi approximation, $|\nabla \rho_0| \ll \rho_0$ and $\nabla \mathbf{v}_0^2 \ll 2n_0$, the coarse-grained density background $\rho_0 = 1-\frac{1}{2}\mathbf{v}^2$ is given by
\begin{equation} \label{dens}
\rho_0 = \begin{cases}
1+\frac{1}{2}\left(\ell/R\right)^2\left(r/R\right)^2-\left(\ell/R\right)^2, 
\quad r\leq R, \\
1-\frac{1}{2}\left(\ell/r\right)^2, \qquad \ \ r>R
\end{cases}
\end{equation}
for $\rho_0 \gtrsim 0$ and $\rho_0 \approx 0$ otherwise. Hence, the solid body region is absent from the flow for $R<\ell/\sqrt{2}$ and the cluster transitions to a single multiply winded vortex (c.f.~Fig.~\ref{fig:sim}), requiring corrections beyond the Thomas-Fermi approximation.

\begin{figure*} 
\centering
\includegraphics[width=\linewidth]{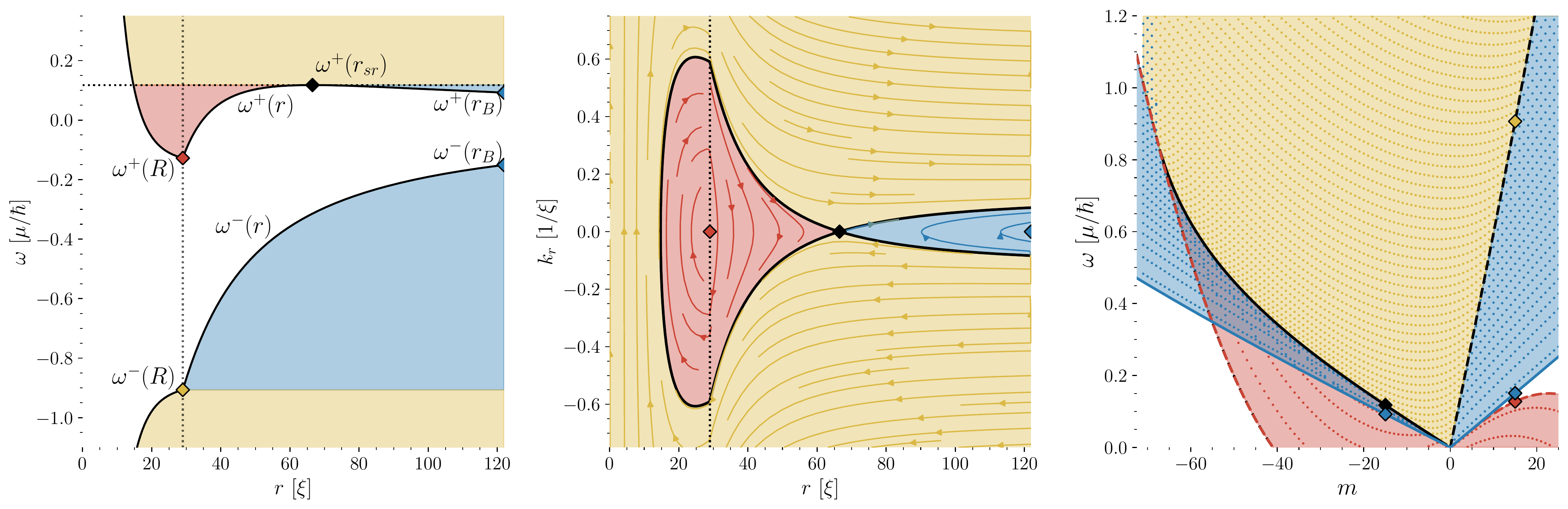}
\caption{\textbf{Analytical predictions.} Three qualitatively different regions shaded: waves bound to the core region (red), waves trapped outside (blue), and waves that communicate between the core and the boundary (yellow). Left panel: the effective potential $\omega^\pm(r)$ (solid black line) for $m=-15$. The symmetry $-\omega_D^\pm(m)=\omega_D^\mp(-m)$ means that negative frequencies inform us of $m=+15$. Middle panel: radial phase space trajectories for waves with azimuthal number $m=-15$. Right panel: allowed frequencies $\omega$ over azimuthal number $m$ for all radii $r$ in the system. For a single $m=\pm 15$, this corresponds to the left panel projected onto the frequency axis.
} \label{fig:model}
\end{figure*}

{\textbf{Perturbative expansion.}}--- To model the frequency content of the sound field, Eq.~\eqref{SGPE} is linearized about the slowly moving background $\partial_t \phi_0, \partial_t \rho_0 \ll 1$, i.e.~$\partial_t R \ll 1$. The resulting dynamics for the fluctuations is given by 
\begin{subequations}
\label{lin1}
\begin{align}
    \left(\partial_t + \mathbf{v}\cdot \nabla\right) \delta \rho &= - \rho_0 \nabla^2 \delta \phi~, \label{lin1a} \\ 
    \left(\partial_t + \mathbf{v}\cdot \nabla\right) \delta \phi &= - \left(1-\frac{\nabla^2}{4\rho_0}\right)\delta \rho ~. \label{lin1b}
\end{align}
\end{subequations}
Here, we neglected $\eta, \gamma \ll 1$, which act as sources and sinks of fluctuations but do not significantly alter the sound-field frequencies. In the long wavelength limit, the excitations are phonons described by a single relativistic Klein-Gordon equation on an effective spacetime determined by the background flow $\mathbf{v}$. This mathematical equivalence builds the basis of analogue gravity systems~\cite{barcelo2011analogue}.

To model the sound-waves, we apply a WKB approximation, which amounts to
assuming that the solutions to \eqref{lin1} look locally like plane waves
\begin{equation} \label{wkb}
\begin{bmatrix}
\delta \phi \\ \delta \rho
\end{bmatrix} = \begin{bmatrix}
\mathcal{A}(r) \\ \mathcal{B}(r)
\end{bmatrix}e^{i\int k_r(r)dr+im\theta-i\omega t},
\end{equation}
with a slowly varying amplitude, i.e. $\partial_r k_r\ll k_r^2$, $\partial_r \mathcal{A}\ll k_r\mathcal{A}$ and similarly for $\mathcal{B}$. Here, we have also used the independence of $t$ and $\theta$ in \eqref{vel} and \eqref{dens} to decompose into separate frequency $\omega$ and azimuthal $m$ components. Inserting this ansatz into \eqref{lin1} and keeping only the dominant terms gives the local dispersion relation,
\begin{equation} \label{disp}
\omega_D^\pm 
= \frac{mv_\theta}{r} \pm \sqrt{\rho_0 k^2+\tfrac{1}{4}k^4}, 
\qquad k = \sqrt{k_r^2+\frac{m^2}{r^2}}
\end{equation}
where a wave which satisfies the equations of motion must lie either on the upper branch ($\omega=\omega_D^+$) or on the lower branch ($\omega=\omega_D^-$).

\begin{figure*} 
\centering
\includegraphics[width=\linewidth]{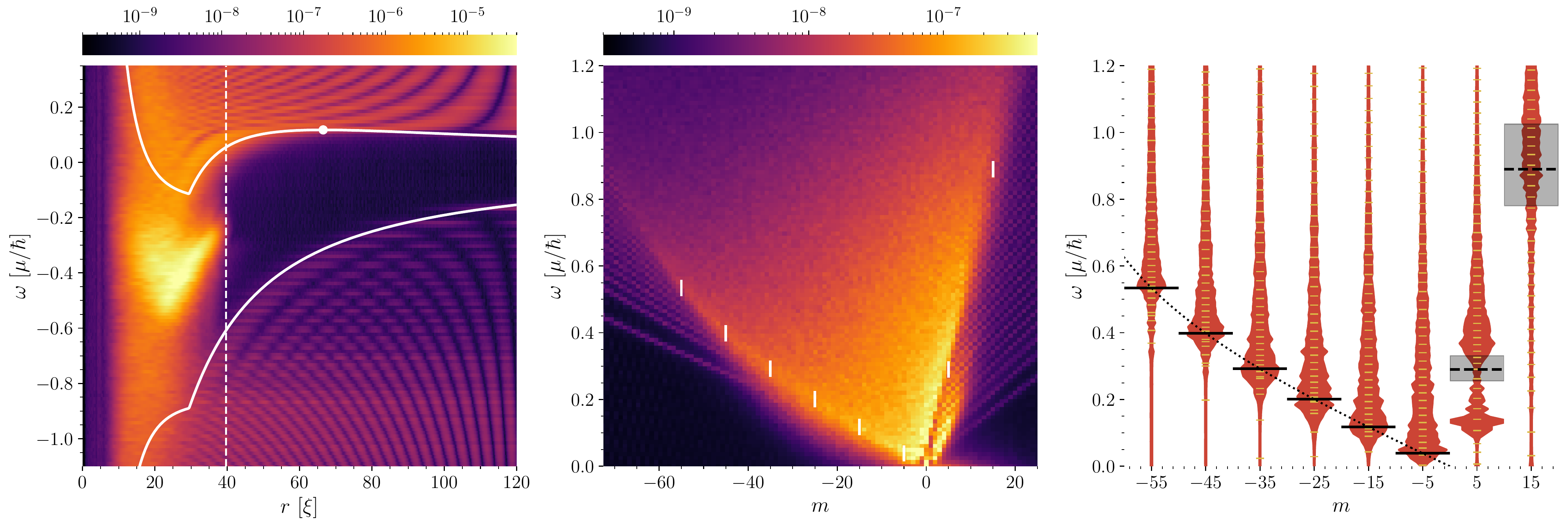}
\caption{\textbf{Simulation results.} 
Left panel: ensemble-averaged squared amplitudes for $m=\pm 15$. Central panel: radially averaged spectrum from $r=40$ (see white vertical dashed line in left panel) to the boundary $r=120$. Right panel: relative amplitudes of central panel at selected values for $m$. Here $\Delta m=10$ corresponds to a amplitude of $0.1$ in that $m$-channel. Black solid horizontal lines are depicting the sound-ring frequencies, and the shaded region for $m>0$ is the delimiting frequency $\omega^+(R)$ from the initial ($R_0=26.2\pm 0.3$) and final ($R_1=32.7\pm 0.5$) ensemble-averaged core size. The black dashed line is the delimiting frequency $\omega^+(r)$ at the average core size $r=R_0/2+R_1/2$. As expected it peaks at the frequencies corresponding to core sizes at earlier times. 
} \label{fig:data}
\end{figure*}

Using \eqref{wkb} we obtain the radial phase-space trajectory $(r(t),k_r(t))$ of a wave that satisfies the dispersion relation $\omega_D^\pm$, by following a trajectory of constant phase~\cite{tracy2014ray}. The resulting Hamiltonian dynamics ${\dot{r}=\partial_{k_r} \omega_D^\pm}$ with ${\dot{k}_r=-\partial_r \omega_D^\pm}$ inform us that waves stagnate radially ($\dot{r}=0$) for $k_r=0$. We refer to the radius $r_\tp$ where a mode stagnates as a turning point, and the frequencies
\begin{equation}
\omega^\pm(r) = \frac{mv_\theta}{r} \pm \sqrt{\rho_0\frac{m^2}{r^2}+\frac{m^4}{4r^4}}
\end{equation}
of modes that stagnate at $r_\tp$, as the turnover frequencies $\omega^\pm$. Note that these curves inform us of the location of the turning points $r_\tp$ for a given frequency $\omega$ through the relation ${\omega=\omega^\pm(r_\tp)}$.
An example of $\omega^\pm(r)$ for ${m=-15}$ is displayed in the left panel of Fig.~\ref{fig:model}. The plotted negative frequencies inform us of the co-rotating ($m>0$) case ${m=15}$ by virtue of the symmetry ${-\omega(m)=\omega(-m)}$ in equation \eqref{disp}.

For positive frequency ($\omega>0$) counter-rotating waves, i.e. waves that rotate in a direction opposite to that of the cluster ($m<0$), the $\omega^+(r)$ frequencies can exhibit a local maximum. This local maximum, which is an unstable fixed-point in radial phase space (see middle panel in Fig.~\ref{fig:model}), is precisely the sound-ring which is situated at,
\begin{equation}
r_\sr = \sqrt{\tfrac{1}{2}(6\ell^2  -m^2)+\ell\sqrt{6\ell^2-m^2}}, \quad \omega_\sr=\omega^+(r_\sr).
\label{eq:soundring}
\end{equation}
Here $r_\sr$ is the radius of the sound-ring and $\omega_\sr$ is the frequency of the mode which orbits the system at this radius, see Fig.~\ref{fig:model}. Since the effective Hamiltonian is locally flat at the sound-ring $r_\sr$, i.e. $\partial_r\omega^+|_{r=r_\sr}=0$, waves which are created in this region will linger around $r_\sr$ before eventually dispersing \cite{Torres_2018}.
Note that the sound-ring only exists when the cluster is sufficiently compact, i.e.~$R<r_\sr$, and only for azimuthal modes with ${|m|\lesssim\sqrt{6}|\ell|}$. For a singly winded vortex ($|\ell|=1$) the sound-ring is of the order of the healing length $r_\sr\sim 1$.

For co-rotating waves $m>0$, for which there is no sound-ring, all frequencies originating in the core propagate freely to the boundary of the condensate. These waves have ${\omega>\omega^+(R)}$ and occupy the yellow region of Fig.~\ref{fig:model}. By contrast, waves with $\omega<\omega^+(R)$ scatter with the barrier $\omega^+$ in the region $r>R$ and therefore do not probe the core. These waves occupy the blue region. 

The existence of a sound-ring for counter-rotating modes alters this picture. As in the previous case, we distinguish between two types of waves which penetrate (yellow)/do not penetrate (blue) the core, with $\omega_\sr$ delimiting the two scenarios, see Fig.~\ref{fig:model}. There are two important differences from the co-rotating case: (1) Waves with $\omega_\sr$ do not travel freely between the core and the boundary. Instead, they linger at the sound-ring radius $r_\sr$ before dispersing both inwards and outwards. (2) There are frequencies below $\omega_\sr$ that exist in the core region, which are prevented from escaping due to the presence of a potential barrier outside the core where the flow is irrotational. This situation, where waves with $\omega<\omega_\sr$ become trapped in a region localised around the core, i.e. $r\sim R$, is shaded red in Fig.~\ref{fig:model}. Notice that for sufficiently compact clusters, the positive branch $\omega_D^+$ of the dispersion relation takes negative values. By virtue of the symmetry $-\omega_D^\pm(-m) = \omega_D^\mp(m)$, these waves appear as co-rotating, positive frequency waves from the lower branch of the dispersion.

In general, a perturbed system will start to settle into a superposition of resonant modes after the initial perturbation has had time to collide with the boundaries of the system. The resonant frequencies are computed used a procedure that is conceptually analogous to computing bound states of the Schr\"odinger equation for a potential that contains either one or two wells within the semi-classical approximation\cite{SM}. The frequencies which solve the resonance conditions are illustrated as points in the right panel of Fig.~\ref{fig:model}.

\textbf{Results.}---
In our simulations, and in experiments, damping is expected to act more aggressively on higher frequencies. Therefore, the majority of the sound-energy escaping the cluster must reside in the frequencies delimiting the yellow and blue regions. Consequently, the peak amplitude for co-rotating frequencies depends on the core size, and therefore on the time the waves were emitted from the core. In comparison, the peak at the sound-ring frequencies remains constant and is thus well resolved. This result is independent of the validity of the model inside the core. 

In Fig.~\ref{fig:data} we display the frequency content of the sound field acquired from our simulations. The left panel illustrates the radial dependence of the sound in different frequencies for $m=-15$. Superimposing the turnover frequencies from our model, we see that the data is in excellent agreement with our prediction for $r>R$, since $\omega^\pm$ correctly separates the propagating and evanescent regions of the system. By contrast, the model fails for $r<R$ since for our choice of $\ell$ (discussed later) waves in the studied $\omega$-range will be sensitive to vortex dynamics. 

Based on this observation, we display the spectrum in the central panel, i.e. the radial average of the modulus squared over $r\in[r_a,r_b]$ of the left panel, for each $m$. We have chosen $r_a=40$ (vertical white dashed line in left panel) to be just outside the largest final core in the ensemble. 
The co-rotating waves are still affected by their dependence on the core size $R$ at the time of departure (see Fig.~\ref{fig:data}). As expected, the counter-rotating power spectrum in the simulations becomes independent of $r_a$ provided $r_a>\mathrm{max}[r_\sr(m)]$. The dominant features in this signal are the sound-ring frequencies and the resonant modes living outside the cluster (corresponding to the blue region in Fig.~\ref{fig:model}). Resonant frequencies in the yellow and red regions are not present since these probe the region $r<R$ where our model breaks down. 

\textbf{Conclusion.} In this work, we introduced the notion of the sound-ring to describe the wave dynamics of a compact vortex.
By direct simulation of the full BEC equations of motion, we then verified that the sound-ring of the average flow field is the dominant signal in the measured spectrum. This complements recent progress in black hole physics, where the gravitational waveform produced during binary mergers is determined by the light-ring of the average one-body spacetime.
In our case, the sound-ring plays a dominant role since it exists in a region of the flow which is highly symmetric and non-evolving over the course of evolution. The resulting signal is therefore independent of the non-linear vortex dynamics inside the cluster, provided the vortices are packed close enough together that the cluster radius is smaller than the sound-ring, i.e. $R<r_{sr}$.

We expect our findings to be of interest in 2d-quantum turbulence. In a confined system such as a trapped atomic condensate, the number of possible vortex configurations is bounded and the entropy reaches a maximum at a finite value of the energy \cite{Onsager}. Larger energy can be reached only by lower entropy creating clustering of vortices of the same sign. Therefore, injection of energy into such a 2d flow will promote the formation of vortex clusters and sound rings. A coarse-grained velocity field of $\ell$ same-sign vortices will induce a sound-ring of radius $r_{sr} \sim \ell$, see Eq.~\eqref{eq:soundring}, in which a dominant wave signal is localised around $\omega_{sr}$ in the frequency domain. This is something which could be checked in current experimental and numerical studies of 2d-quantum turbulence in which vortex clustering can be induced by evaporative heating, i.e.~the annihilation of vortex-antivortex pairs \cite{Johnstone}.

Finally, although in our case we do not observe a discrete spectrum of modes which probe the inside of the cluster, we expect these modes to be excited for larger vortex clusters with $\ell\gtrsim 100$. In this limit, we believe the discretization of angular momentum in the underlying micro-structure will be hidden from the long wavelength modes, which perceive only the average ``classical'' Rankine background. These considerations may be of interest for the corresponding quantum to classical limit around both fluid flows as well as rotating compact objects exhibiting discrete angular momentum. 

\textbf{Acknowledgements.} 
We thank Steffen Biermann, Thiago Cardoso, Cisco Gooding, Maur\'{i}cio Richartz, J\"org Schmiedmayer, Vitor B. Silveira, and William G. Unruh for stimulating discussions. 
CB and SW acknowledge support provided by the Science and Technology Facilities Council on Quantum Simulators for Fundamental Physics (ST/T006900/1) as part of the Quantum Technologies for Fundamental Physics programme.
SW acknowledges support provided by the Leverhulme Research Leadership
Award (RL-2019 - 020), the Royal Society University
Research Fellowship (UF120112) and the Royal Society
Enhancement Grant (RGF/EA/180286), and partial support by the Science and Technology Facilities Council (Theory Consolidated Grant ST/P000703/1).
AG and SW acknowledge support provided by the Royal Society Enhancement Grant (RGF/EA/181015).
SE and SW acknowledge support from the EPSRC Project Grant
(EP/P00637X/1).
SE acknowledges support
through the Wiener Wissenschafts- und TechnologieFonds (WWTF) project No MA16 - 066 (``SEQUEX''),
and SE funding from the European Union's Horizon 2020
research and innovation programme under the Marie
Sklodowska-Curie grant agreement No 801110 and the
Austrian Federal Ministry of Education, Science and Research (BMBWF) from an ESQ fellowship.
SP acknowledges support from the Natural Science and Engineering Research Council of Canada.

\bibliography{bibfile.bib}
\bibliographystyle{apsrev4-2}



\clearpage
\widetext
\setcounter{equation}{0}
\setcounter{figure}{0}
\setcounter{table}{0}
\setcounter{page}{1}

\renewcommand\thesection{Supplementary \Alph{section}}
\renewcommand\thesubsection{\Alph{section}\arabic{subsection}} 
\renewcommand{\theequation}{S\arabic{equation}}
\renewcommand{\thefigure}{S\arabic{figure}}

\begin{appendices}
\section{Supplemental Material}

\subsection{Numerical simulations} \label{supplementary:numerics}
\noindent
We model our system with the stochastic Gross-Pitaevskii equation (SGPE), which in our dimensionless units (see main text) takes the form
\begin{equation}
    i \partial_t \Psi = (1-i\gamma) \left[-\frac{1}{2}\nabla^2 + U-1 + |\Psi|^2\right]\Psi + \varepsilon\eta ~.
\end{equation}
Here, the noise $\varepsilon \eta$ is written in terms of the dimensionless variance $\varepsilon^2 \in \mathbb{R}$ and a Gaussian white noise term $\eta$, sampled from a standard complex normal distribution. The SGPE describes the dynamics of the coherent region $\Psi$ while being in contact with a thermal bath. The dissipative term $\gamma$ represents the particle transfer between the coherent region (low-energy modes) and the thermal reservoir (high-energy modes) and the noise $\eta$ represents the random nature of incoherent scattering within the system. This leads to the system approaching a thermal equilibrium state in the long time limit, with the temperature $T$ determined by the relation $\varepsilon^2 \sim T \gamma$. Note that in our adimensionalized variables, the temperature varies with the atomic species constituting the condensate. For the results presented in the main text, we choose $\gamma=2.5\cdot10^{-3}$ and $\varepsilon^2=2.5\cdot 10^{-3}$. Provided they are reasonably small, the values for the noise $\eta$ and damping $\gamma$ play little role for the results presented in the main text.  However, for the considered system, their presence plays a delicate role in the numerical implementation. 

The damping $\gamma>0$, which introduces energy loss to the system, makes the vortex cluster decay faster and adds stability to the evolution. This greatly reduces the computational cost of a single simulation. Moreover, as a cluster decays into individual vortices, the energy is released as waves. In the absence of damping ($\gamma=0$), one would have to wait a very long time for the cluster to decay to sizes $R\simeq \ell$, and worse, by then the condensate, and its resonances, would be highly populated by the waves emitted in the process. Note, however, that choosing $\gamma$ too large results in overcritical damping of waves and makes the cluster decay too fast for the sound-ring phenomena to be relevant. 

The main role of the noise $\varepsilon \eta$ is to generate realistic vortex configurations in the early stage of the relaxation by explicitly breaking the rotational symmetry of the initial state. To avoid imposing a specific vortex configuration, the initial state of the simulation is a multiply winded central vortex (defined precisely later). For such a state to decay into a cluster of individual vortices, there must be some feature in the system that provides the seeds for breaking the symmetry of the initial state. In the absence of noise, the only non-rotationally symmetric feature in the simulation is the cartesian grid, which results in early vortex configurations inheriting the $\mathbb{Z}_4$ symmetries of the discretization. Inclusion of noise, as e.g.~quantum or the considered thermal fluctuations, alleviates this problem and leads to realistic random vortex configurations. 

The numerical discretization $\Psi_{n,i,j} \equiv \Psi(t_n,x_i,y_j)$ of the wave-function $\Psi$ is done onto a set of equal-sized cartesian voxels
\begin{equation*}
    \{t_n\}\times\{x_i\}\times\{y_j\}\subseteq\mathbb{R}^{N_t}\times\mathbb{R}^{1536}\times\mathbb{R}^{1536} ~.
\end{equation*}
To resolve individual vortices, a spatial grid-spacing $\Delta x \equiv x_{i+1}-x_{i}=1/5$ is chosen. The time-step $\Psi_{n,i,j}$ to $\Psi_{n+1,i,j}$ is performed using a Fourier-spectral Strang-splitting scheme as outlined in \cite{javanainen2006symbolic}. The contribution from noise $\eta$ is added after each timestep using $\Psi_{n+1,i,j} \mapsto \Psi_{n+1,i,j} + \varepsilon\eta\Delta t$. Although the stability of the solver varies with the damping $\gamma$, the presented simulations are using $\Delta t = 5 \cdot 10^{-3}$, which is well within the stable region. To enforce periodicity of $\Psi$ at the boundaries $x,y \in \{\pm 1536 \Delta x/2\}$ of the simulation domain, and thus the adequacy of the Fourier-spectral scheme, a sharp logistic trap of the form $U(r)\simeq 10/(1+\exp(2(r_B-r)))$ with $r_B \simeq 614\Delta x$ is used. 

Each simulation is initialized in the ground state of the external potential $U(r)$ by evolving a uniform condensate under the influence of an initial strong damping (imaginary time evolution). Once completed, the noise and damping are set to their final small values and we imprint a multiply winded vortex with $\ell = 29$ in the center of the potential. Here we use the approximate expression
\begin{equation}
    \Psi_{n,i,j} \mapsto \Psi_{n,i,j} e^{i\ell \theta(x_i,y_j)}
\end{equation}
with the density adjusted according to \eqref{dens} with a flat ($\rho_0=0$) core inside $r=\ell/\sqrt{2}$. Note that due to the existence of damping the vortex core relaxes to its exact form long before the cluster decays into individual vortices.

\subsection{Post-processing and data analysis} 
\begin{figure}
    \centering
    \includegraphics[width=\textwidth]{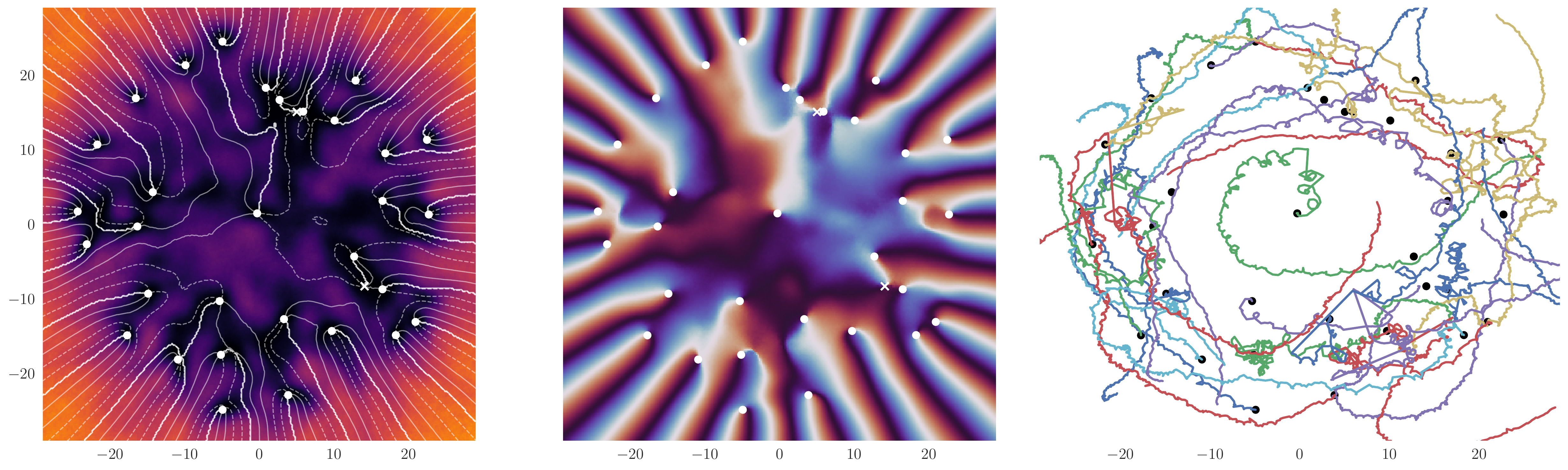}
    \caption{
    \textbf{Vortex tracking results.} Left panel: density of the first time frame in in the analysed window of a single realisation. White lines are drawn along curves of constant phase. White circular dots are the detected phase singularities, white crosses are the locations of the phase singularities with inverted winding with respect to the net rotation of the cluster. 
    Middle panel: phase information of the data shown in the left panel. The dots and crosses are the same as in the left panel. 
    Right panel: traced trajectories of phase singularities (vortices) in the co-rotating frame. Black dots are the initial phase singularities shown in the left and middle panel.
    }
    \label{fig:vortex-tracking-triplet}
\end{figure}

\noindent
Expectation values are calculated averaging over 51 independent realizations $\{\Psi_a(t_n,x_i,y_j)\}$ referred to as an ensemble. Each $\Psi_a$ is cropped around a time window with $512$ time-frames $t\in[t_1,t_2]$ with $\Delta t=1$, such that the wavefunction has time-averaged cluster radius $R=(1\pm0.02)\ell$. The resulting ensemble averaged cluster size in the window is $R\in[26.2\pm 0.3,32.7\pm 0.5]$.
Here, $R$ is estimated by fitting the wavefunction to the course-grained velocity field in equation~\eqref{vel}. To make sure this fit is adequate, it was tested against vortex-tracking results and similar fits of the densities $\rho^{(a)}=|\Psi_a|^2$ to \eqref{dens}. The result of such a vortex tracking procedure is exhibited in Fig.~\ref{fig:vortex-tracking-triplet}. The vortex trajectories in the right panel are plotted in the frame co-rotating with the solid body flow in the core. Two features should be noticed in this figure: (1) The vortex density is slightly larger in the outer regions of the core. The consequence of this is a modification of the solid body flow inside the core, wherein the peripheral region rotates faster than the central region. (2) Even in the co-rotating frame, the vortex trajectories are seen to be very chaotic. 

The density data $\rho^{(a)}$ in the time-window is interpolated onto a polar mesh and decomposed into the different frequency and azimuthal components using a temporal Hilbert transform. The result is a complex function
\begin{equation}
\rho_m^{(a)}(\omega,r) = \int^{t_2}_{t_1} \int^{2\pi}_0\rho^{(k)}(t,r,\theta) e^{-im\theta+i\omega t} d\theta\,dt.
\label{eq:numericsDensity}
\end{equation}
Finally, we average over a radial window, and over the ensemble, to obtain the spectrum of the cluster $\langle |\rho_m^{(a)}(\omega)|^2\rangle _{r,a}$. This is the quantity plotted in the central panel of Fig.~\ref{fig:data}. The quantity plotted in the left panel is the ensemble-averaged \eqref{eq:numericsDensity} for a given $m$, i.e. $\langle |\rho_m^{(a)}(\omega,r)|^2\rangle _{a}$.

\subsection{The Velocity field of a quantized vortex cluster} 
\label{supplementary:vortex-theory}
\begin{figure}
    \centering
    \includegraphics[width=\textwidth]{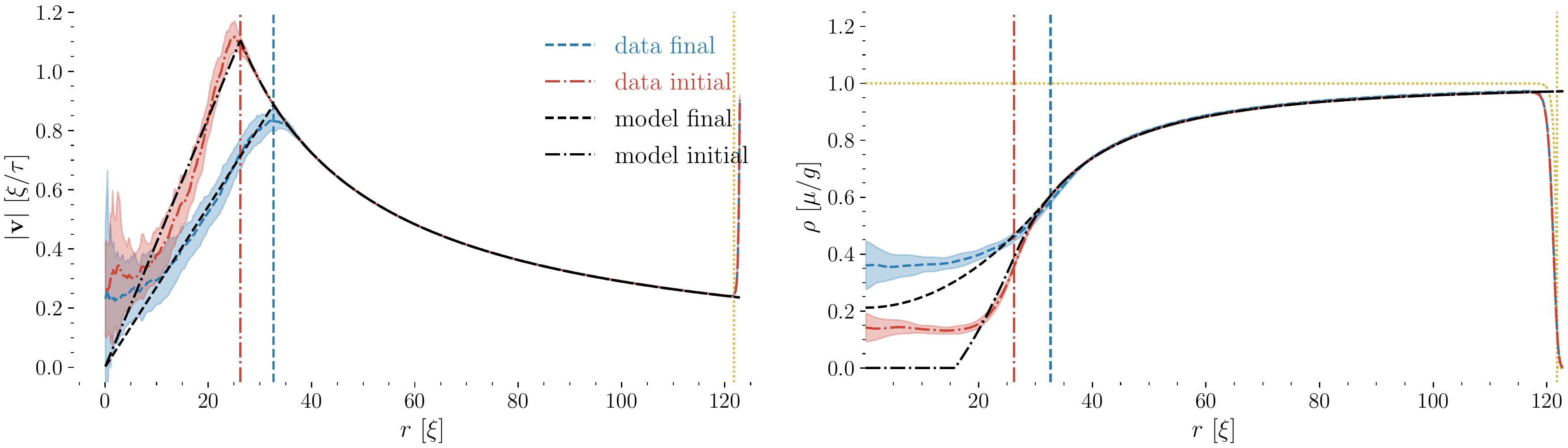}
    \caption{
    \textbf{Measured background.} Left panel: measured velocities $\mathbf{v}=\text{Im}\nabla \ln \Psi_a$ at the initial (red dot-dashed) and final (blue dashed) timeframe. The ensemble averaged azimuthal mean is plotted with the standard deviation over ensemble shaded. Right panel: azimuthally averaged density profile $\rho^{(a)}(r)$ at the initial (red dot-dashed) and final (blue dashed) timeframe. The density is compared to the uniform Thomas-Fermi density $\sqrt{1-U(r)}$ (yellow dotted). 
    In both panels, the model is fitted to find the core size $R$ (vertical lines) and plotted as black lines on top.
    }
    \label{fig:densityAndVelocity}
\end{figure}

\noindent
For a collection of $N$ vortices at positions $\{(x_k,y_k)\} \subset \mathbb{R}^2$ with windings $\{\ell_k\} \subset \mathbb{Z}\setminus\{0\}$, the background velocity potential may be approximated as the sum of individual (boundary-independent) windings provided that they are well over a healing length apart,
\begin{equation}
  \phi_0 = \sum_{k=1}^N \ell_k\arctan \left(\frac{y-y_k}{x-x_k}\right).
  \label{eq:effectivePhi}
\end{equation}
The gradient $\mathbf{v}\equiv \nabla \phi_0$ can be conveniently written in terms of complex variables $z = x+iy$ as
\begin{equation}
  v^* = \sum_{k=1}^N  \frac{i \ell_k}{z_k-z}
  \label{eq:velocityfield}
\end{equation}
where $*$ denotes the complex conjugate. Now focus on the case where the center of vorticity is in the origin, i.e. $\sum_n \ell_k z_k = 0$, and impose an ordering $0\leq |z_1| \leq ... |z_M| \leq |z| \leq |z_{M+1}| ... \leq |z_N| \equiv R$. Introduce polar variables $z=re^{i\theta}$ and denote the azimuthal average by $\langle \cdot \rangle$, then it follows from the residue theorem that 
\begin{equation}
  \langle ve^{-i\theta} \rangle^* 
   = \sum_{k=1}^N  \frac{\ell_k}{2\pi r}\oint_{|z|=r} \frac{dz}{z_k-z} 
   = -\frac{i}{r} \sum_{k=1}^M  \ell_k.
\end{equation}
so that
\begin{equation}
  \langle \mathbf{v} \rangle = \frac{\ell_M}{r}\mathbf{e}_\theta 
  ~ \text{ for } ~ \ell_M \equiv \sum_{k=1}^M  \ell_k ~,
  \label{eq:averagedVelocity0}
\end{equation}
where $\mathbf{e}_\theta \equiv ie^{i\theta}$. That is, the azimuthally averaged velocity field is that of a single, central vortex with a winding equal to the sum of the windings of the vortices contained inside the circle of radius $|z|$. In fact, outside the cluster $|z|>|z_N|$, this relation holds to second order in $|z_N/z|$ without the need for an azimuthal average
\begin{equation}
\mathbf{v} = \frac{\ell}{r}\mathbf{e}_\theta + \mathcal{O}\left(\left|\frac{z_N}{z}\right|^2\right) ~ \text{ for } ~ \ell \equiv \sum_{k=1}^N \ell_k.
\label{eq:vOutside}
\end{equation}
Assuming a uniform distribution of vortices inside $R$, i.e. $\mathrm{d}\ell \propto \mathrm{d}A$, we may write
\begin{equation}
    \ell_M \simeq \ell \left(\frac{r}{r_c}\right)^2 
    \text{ where } 
    r_c \equiv \begin{cases}
    & R  \text{ for } r\leq R \\
    & r  \text{ for } r > R 
  \end{cases}
  \label{eq:RankineRadius}
\end{equation}
equation \eqref{eq:RankineRadius} takes the form of a drain-free Rankine vortex $\mathbf{v} = v_\theta(r) \mathbf{e}_\theta$ with
\begin{equation}
  v_\theta = 
  \frac{\ell}{r}\left(\frac{r}{r_c}\right)^2 = 
  \begin{cases}
    & \frac{\ell r}{R^2}  \text{ for } r\leq R \\
    & \frac{\ell }{r}  \text{ for } r > R 
  \end{cases}
  \label{Seq:rankineVelocity}
\end{equation}
To find an expression for the background solution $\rho_0=1-\frac{1}{2}\mathbf{v}^2$ we need to obtain an expression for $\langle \mathbf{v}^2 \rangle$ inside the cluster. Note that the need for this calculation arises because the azimuthally averaged velocity field is no longer irrotational inside the cluster. The following procedure is equivalent to the result one would obtain from including the integral of $\mathbf{v}\times (\nabla \times \mathbf{v})$ from $r$ to infinity in the Bernoulli equation. The outside is already given from equation \eqref{eq:vOutside}, but to find the value inside the cluster we observe that
\begin{align*}
    v^*v  
    &= \sum_{m,n} \frac{\ell_n \ell_m }{(z_n-z)(z_m^*-z^*)}
    \simeq 
    r^6\sum_{m,n} \ell_n \ell_m 
    \left[ \frac{\mathbb{I}_{n\leq M}}{r^4} - \frac{\mathbb{I}_{n>M}}{(z_n z^*)^2} \right]
    \left[ \frac{\mathbb{I}_{m\leq M}}{r^4} - \frac{\mathbb{I}_{m>M}}{(z_m^* z)^2} \right] \\
    & =
    \left[ \frac{\ell_M}{r} - \sum_{n>M}\frac{r^3\ell_n}{(z_n z^*)^2} \right]
    \left[ \frac{\ell_M}{r} - \sum_{n>M}\frac{r^3\ell_n}{(z_n^* z)^2} \right] 
    =
    \frac{\ell_M^2}{r^2} 
    -\frac{\ell_M}{r^2}\sum_{n>M}\ell_n\left[
    \left(\frac{z}{z_n}\right)^2+\left(\frac{z^*}{z_n^*}\right)^2
    \right] 
    + 
    \sum_{n>M} \sum_{m>M}
    \frac{r^2\ell_m \ell_n}{(z_n z_m^*)^2} ~,
\end{align*}
where $\mathbb{I}_{C}$ is the indicator function. Taking the azimuthal average removes the entire middle term and all off-diagonal entries in the last term, leaving only
\begin{equation}
    \langle v^* v \rangle
    \simeq
    \frac{\ell_M^2}{r^2} + 2\pi r^2 \sum_{n>M} \frac{\ell_n^2 }{r_n^4 }
\end{equation}
For a uniform vortex distribution, only the diagonal terms $n=m$ in the sum contributes, leaving us with
\begin{equation}
    \langle |\mathbf{v}|^2 \rangle 
    \simeq
    \frac{\ell^2}{r_c^2}\left[2-\frac{r^2}{r_c^2}\right]
    = 
    \begin{cases}
    \frac{\ell^2}{R^2}\left[2-\frac{r^2}{R^2}\right] & \text{ for } r\leq R \\
    \frac{\ell^2}{r^2} & \text{ for } r> R \\
    \end{cases}
    \label{eq:vSquared}
\end{equation}
This finally allows expressing the background density 
\begin{equation}
    \rho_0(r) = 1-\frac{1}{2}\frac{\ell^2}{r_c^2}\left[2-\frac{r^2}{r_c^2}\right].
    \label{Seq:backgroundDensity}
\end{equation}
The models for the background density $\rho_0(r)$ in equation \ref{Seq:backgroundDensity} and velocity $v_\theta(r)$ in \ref{Seq:rankineVelocity} is found to be in good agreement with numerical results (see Fig.~\ref{fig:densityAndVelocity}).

\subsection{Computing the resonant modes}
\label{supplementary:resonantModes}
Since the system is spatially finite, our model may populate resonant modes.
We compute these by imposing the boundary conditions and accounting for the mode specific scattering induced by the effective potential $\omega^\pm(r)$.
This procedure is conceptually analogous to computing bound states of the Schr\"odinger equation for a potential that contains either one or two wells within the semi-classical approximation. 

To leading order, the WKB expansion of the perturbative dynamics \eqref{lin1} gives the dispersion relation \eqref{disp}. The constraint imposed by the dispersion relation can be thought of as the requirement that the function
\begin{equation*}
    H \equiv 
    \frac{1}{2}\left(\omega-\frac{mv_\theta}{r}\right)^2
    -\frac{1}{2}k^2\left(\rho_0+\frac{1}{4}k^2\right), \ k^2 \equiv k_r^2 + \frac{m^2}{r^2}
\end{equation*}
is zero, i.e. $H=0$. This is exactly a Hamilton-Jacobi equation with the phase $\text{Im}( \ln \delta \rho)$ acting as the principal/action function. The WKB propagation of waves can then be interpreted as the flow in phase space generated by the Hamiltonian $H$. While the rotational invariance of $H$ allow us to study the reduced, radial phase $(r,k_r)$ for each $m$, problems arise on turning points $\partial_{k_r}H = 0$. At these points, the WKB assumption of a phase varying much faster than the amplitude is violated. In this section, we search for solutions to the wave equation close to a turning point, and use the asymptotic expansions of these solutions to relate WKB solutions on opposite sides of the turning point. 

To start, we consider a turning point $(r_0,{k_r}_0)$ and expand the Hamiltonian around this point
\begin{equation}
    \tilde{H} \equiv
    (\partial_r H)(r-r_0) + \frac{1}{2}(\partial_{k_r}^2H)(k_r - {k_r}_0)^2 ~,
    \label{eq:HamiltonianExpansionTurningPoint0}
\end{equation}
where it is understood that $\partial_r H$ and $\partial_{k_r}^2H$ is evaluated at the turning point. Next, we relax the WKB assumption by promoting $k_r$ to a differential operator $-i\partial_r$. The result is a Hamiltonian operator
\begin{align}
    \mathcal{H} 
    &=
    (\partial_r H) (r-r_0) + \frac{1}{2}(\partial_{k_r}^2H) ({k_r}_0^2 + 2{k_r}_0 i\partial_r - \partial_r^2)
    \label{eq:HamiltonianExpansionTurningPoint}
\end{align}
determining the solutions $\phi$ at the turning point through the differential equation $\mathcal{H}\phi = 0$. If $\partial_{k_r}^2H\neq 0$ at the turning point, then the equation $\mathcal{H}\phi = 0$ takes the form
\begin{align}
    -\partial_r^2 \phi
    +2i{k_r}_0\partial_r \phi
    +\left[{k_r}_0^2+Q(r-r_0)\right]\phi = 0
\end{align}
for $Q \equiv 2(\partial_r H)/(\partial_{k_r}^2 H)$. Factoring out a plane wave with $\phi \equiv \psi\exp(i{k_r}_0 r)$, and substituting $z \equiv Q^{1/3}(r-r_0)$ this can be brought to the form of an \textit{Airy equation} $\psi''(z) = z\psi(z)$, whose solutions are Airy functions $Ai(z)$ and $Bi(z)$. Therefore, the solution to the wave equation around a turning point $(r_0,{k_r}_0)$ is 
\begin{equation}
    \phi = e^{i{k_r}_0 r}\left[C_1 Ai(z)+C_2 Bi(z)\right]
    \label{eq:tpSolutionExact}
\end{equation}
for two constants $C_1$ and $C_2$. For the purpose of matching the WKB modes on either side of the turning point, the only thing we need from the Airy functions is their asymptotic expansion:
\begin{subequations}
\label{eq:AiryAsymptotic}
\begin{align}
    Ai(z) & \simeq \frac{|z|^{-\frac{1}{4}}}{2\sqrt{\pi}} \begin{cases}
    e^{-\frac{2}{3}z^\frac{3}{2} }
    & \text{ for } z\rightarrow \infty \\
    2\cos\left(-\frac{2}{3}(-z)^\frac{3}{2}+\frac{\pi}{4}\right) 
    & \text{ for } z\rightarrow -\infty
    \end{cases} \label{eq:AiAsymptotic}\\
    Bi(z) & \simeq \frac{|z|^{-\frac{1}{4}}}{2\sqrt{\pi}} \begin{cases}
    2e^{\frac{2}{3}z^\frac{3}{2} }
    & \text{ for } z\rightarrow \infty \\
    2\sin\left(-\frac{2}{3}(-z)^\frac{3}{2}+\frac{\pi}{4}\right)
    & \text{ for } z\rightarrow -\infty
    \end{cases} \label{eq:BiAsymptotic}
\end{align}
\end{subequations}
Before matching with the WKB solutions close to this point, we first note that \eqref{eq:HamiltonianExpansionTurningPoint0} can be inverted to inform us that if $\tilde{H}=0$ then
\begin{equation*}
     \int k_r dr
     = 
     {k_r}_0r \pm \frac{2}{3}(-z)^\frac{3}{2} + \text{const}
\end{equation*}
Therefore, we may match WKB modes of the form $\mathcal{A}\exp(i\int k_r dr)$ with the asymptotic expansions of \eqref{eq:tpSolutionExact} independent of the evolution of the amplitude. Matching with both $z \sim -\infty$ and $z\sim\infty$ results in the linear relation
\begin{align*}
    & 
    \left[\begin{matrix}
    A_+ \\ A_- 
    \end{matrix}\right]
    =
    M
    \left[\begin{matrix} B_+ \\ B_- \end{matrix}\right]
    \text{ where } 
    M \equiv e^{-\frac{\pi i}{4}}\frac{1}{2}
    \left[\begin{matrix}
    i & 2 \\  
    1 & 2i
    \end{matrix}\right]
\end{align*}
relating amplitudes defined by 
\begin{equation*}
    B_+ e^{\frac{2}{3}z^\frac{3}{2} } 
    + B_- e^{-\frac{2}{3}z^\frac{3}{2} }
    \overset{M}{\mapsto}
    A_+ e^{\frac{2}{3}i(-z)^\frac{3}{2}} 
    + A_- e^{-\frac{2}{3}i(-z)^\frac{3}{2}}.
\end{equation*}
Now, we introduce naming conventions where $k_r^{\pm}$ refers to propagating modes with positive ($+$) and negative ($-$) radial wavenumber $k_r$. Likewise, $\tilde{k}_r^\pm$ refers to the evanescent modes that grow ($-$) and decay ($+$) with radius $r$. To match, we note that $z<0$ whenever $Q>0$ and $r<r_0$, or $Q<0$ and $r>r_0$. Then, for $Q>0$ we may associate $A_\mp$ above with the propagating WKB modes $k_r^\pm$, and $B_\pm$ with the evanescent WKB modes $\tilde{k}_r^{\mp}$ (notice the different sign). For $Q<0$, the direction of $z$ is opposite to that of $r$ so $M$ maps inner amplitudes to outer amplitudes, and $M^{-1}$ maps outer amplitudes to inner amplitudes. That is, $M^{-1}$ maps outer amplitudes $A_\pm$ corresponding to $k_r^\pm$ into inner amplitudes $B_\pm$ corresponding to $\tilde{k}_r^\pm$.

Staying with this notation, we may depict the transition as a graphical mnemonic (with radius growing towards the right) 
\begin{align*}
&\begin{tikzpicture}
    \begin{feynman}
    \vertex at (0,0) (v) [blob,fill=white] {$M$};
    \vertex at (1.5,1)  (a1) {$\tilde{k}_r^-$};
    \vertex at (1.5,-1) (a2) {$\tilde{k}_r^+$};
    \vertex at (-1.5,1) (b1) {$k_r^-$};
    \vertex at (-1.5,-1)(b2) {$k_r^+$};
    \diagram* {
      (b2) -- [fermion] (v) -- [fermion] (b1);
      (a1) -- [charged boson] (v) -- [charged boson] (a2);
    };
  \end{feynman}
\end{tikzpicture} 
&\begin{tikzpicture}
    \begin{feynman}
    \vertex at (0,0) (v) [blob,fill=white] {$M^{-1}$};
    \vertex at (1.5,1)  (a1) {$k_r^+$};
    \vertex at (1.5,-1) (a2) {$k_r^-$};
    \vertex at (-1.5,1) (b1) {$\tilde{k}_r^+$};
    \vertex at (-1.5,-1)(b2) {$\tilde{k}_r^-$};
    \diagram* {
      (a2) -- [fermion] (v) -- [fermion] (a1);
      (b1) -- [charged boson] (v) -- [charged boson] (b2);
    };
  \end{feynman}
\end{tikzpicture}
\\ 
& M \equiv \frac{1}{2} e^{-\frac{\pi i}{4}}
    \left[\begin{matrix}
    i & 2 \\  
    1 & 2i
    \end{matrix}\right]
& M^{-1} \equiv \frac{1}{2} e^{-\frac{\pi i}{4}}
    \left[\begin{matrix}
    2 & 2i \\  
    i & 1
    \end{matrix}\right]    
\end{align*}
for $Q>0$ and $Q<0$ respectively. Here, arrows point to the right if the phase, real or imaginary part, increases with radius, and to the left if it decreases. 
That is, arrows on propagating lines indicate the direction of increasing phase with $r$, i.e. the direction of propagation of a positive frequency (and positive branch of the dispersion relation) modes. The arrows on the evanescent modes indicate the direction in which the mode decays. 

We note that if we define the flipping matrix $F$ by $F(a,b)=(b,a)$ then $MF$ flips the right arms of the diagram (swaps column vectors of $M$), and $FM$ swaps the left arms of the diagram (swaps row vectors of $M$). It is therefore convenient to define the matrix $T\equiv FMF$ such that the diagrams take on consistent ``arrow'' conventions: 
\begin{align*}
&\begin{tikzpicture}
    \begin{feynman}
    \vertex at (0,0) (v) [blob,fill=white] {$T$};
    \vertex at (1.5,1)  (a1) {$\tilde{k}_r^+$};
    \vertex at (1.5,-1) (a2) {$\tilde{k}_r^-$};
    \vertex at (-1.5,1) (b1) {$k_r^+$};
    \vertex at (-1.5,-1)(b2) {$k_r^-$};
    \diagram* {
      (b1) -- [fermion] (v) -- [fermion] (b2);
      (a2) -- [charged boson] (v) -- [charged boson] (a1);
    };
  \end{feynman}
\end{tikzpicture} 
&\begin{tikzpicture}
    \begin{feynman}
    \vertex at (0,0) (v) [blob,fill=white] {$\tilde{T}$};
    \vertex at (1.5,1)  (a1) {$k_r^+$};
    \vertex at (1.5,-1) (a2) {$k_r^-$};
    \vertex at (-1.5,1) (b1) {$\tilde{k}_r^+$};
    \vertex at (-1.5,-1)(b2) {$\tilde{k}_r^-$};
    \diagram* {
      (a2) -- [fermion] (v) -- [fermion] (a1);
      (b1) -- [charged boson] (v) -- [charged boson] (b2);
    };
  \end{feynman}
\end{tikzpicture}
\\ 
& T \equiv FMF = \frac{1}{2} e^{-\frac{\pi i}{4}}
    \left[\begin{matrix}
    2i & 1 \\  
    2  & i
    \end{matrix}\right]
&   \tilde{T} \equiv M^{-1} = \frac{1}{2} e^{-\frac{\pi i}{4}}
    \left[\begin{matrix}
    2 & 2i \\  
    i & 1
    \end{matrix}\right]   
\end{align*}
Modes with frequencies larger than the sound-ring frequency (yellow modes) are then determined by the following diagram
\begin{equation}
\begin{tikzpicture}
  \begin{feynman}
    \vertex at (-1,1) (b1) [crossed dot,color=black]{};
    \vertex at (-1,-1) (b2) ;
    \vertex at (1,0) (v) [blob,fill=white] {$\tilde{T}$};
    \vertex at (3,0)(B) [blob,fill=white] {$B$};
    \diagram* {
      (b1) -- [charged boson] (v) -- [charged boson] (b2),
      (B) -- [fermion,quarter left] (v) -- [fermion,quarter left] (B);
    };
  \end{feynman}
\end{tikzpicture}
\label{fig:AmodeDiagram}
\end{equation}
where the crossed circle $\otimes$ indicates an amplitude that is zero. Here, the $B$ refers to the boundary condition. We shall consider an energy-conserving boundary condition of the form $A_- = e^{2i\eta} A_+$. In our case, the WKB propagator from radius $r_i$ to $r_j$ is given by
\begin{align*}
    \mathcal{W}_{ji}
    \equiv 
    Q_{ji}
    \left[\begin{matrix}
    e^{+i\Phi_{ji}}  &   0                   \\
    0                   &   e^{-i\Phi_{ji}}
    \end{matrix}\right]
    \text{ for } \Phi_{ji} \equiv \int_{r_i}^{r_j} k_r dr \\
\end{align*}
where $Q_{ij}$ is a geometrical amplitude factor satisfying $Q_{ij}Q_{ji}=1$. The equation corresponding to diagram \eqref{fig:AmodeDiagram} then takes the form
\begin{equation*}
    \left[\begin{matrix} 0 \\ \tilde{b}_- \end{matrix}\right]
    = \tilde{T}\mathcal{W}_{IB} A_+ \left[\begin{matrix} 1 \\ e^{2i\eta} \end{matrix}\right],
\end{equation*}
where $r_I$ is the turning point inside the cluster core, and $r_B$ is the outer boundary of the condensate. For this to permit non-trivial solutions, we must have 
\begin{equation}
    \cos\left(\eta + \frac{\pi}{4} + \int_{r_I}^{r_B} k_r dr\right) = 0 ~.
    \label{eq:AmodeCondition}
\end{equation}
This is the resonance condition for the modes that may propagate freely between the core and the boundary -- the yellow region in Fig.~2. 

For frequencies smaller than the sound-ring frequency, the wave needs to tunnel through the effective potential $\omega_\pm(r)$. This happens due to the introduction of a new pair of turning points $r_1$ and $r_2$. That is, the diagram for the corresponding modes is
\begin{equation}
\begin{tikzpicture}
  \begin{feynman}
    \vertex at (-1,0) (v) [blob,fill=white] {$\tilde{T}$};
    \vertex at (1,0) (V1) [blob,fill=white] {$T$};
    \vertex at (-3,1) (b1) [crossed dot,color=black]{};
    \vertex at (-3,-1) (b2) ;
    \vertex at (3,0) (V2) [blob,fill=white] {$\tilde{T}$};
    \vertex at (5,0)(B) [blob,fill=white] {$B$};
    \diagram* {
      (v)  -- [fermion,quarter left] (V1) -- [fermion,quarter left](v),
      (b1) -- [charged boson] (v) -- [charged boson] (b2),
      (V1) -- [charged boson,quarter left] (V2) -- [charged boson,quarter left] (V1);
      (B) -- [fermion,quarter left] (V2) -- [fermion,quarter left] (B);
    };
  \end{feynman}
\end{tikzpicture}
\label{fig:IOmodeDiagram}
\end{equation}
whose equation is given by
\begin{equation*}
    \left[\begin{matrix} 0 \\ \tilde{b}_- \end{matrix}\right]
    = \tilde{T}\mathcal{W}_{I1} T  \mathcal{W}_{J2} \tilde{T} \mathcal{W}_{2B} A_+ \left[\begin{matrix} 1 \\ e^{2i\eta} \end{matrix}\right]
\end{equation*}
The result is
\begin{equation} 
    4\cot \left(\Phi_{1I}\right) \cot \left(\eta+\frac{\pi}{4}+\Phi_{B2}\right)
    =
    \exp \left(-2\int_{r_1}^{r_2}|k_r|dr\right)
    \label{eq:IOmodeCondition}
\end{equation}
For a severe evanescent separation $|k_r(r_2-r_1)|\gg 1$ this condition splits into two separate conditions 
\begin{subequations}
\label{eq:IOmodeConditions}
\begin{align}
    & \cos\left(\int_{r_I}^{r_1} k_r dr\right)=0 \label{eq:ImodeCondition}
    \\ 
    & \cos\left(\eta+\frac{\pi}{4}+\int_{r_2}^{r_B}k_r dr\right)=0 \label{eq:OmodeCondition}
\end{align}
\end{subequations}
which for a Neumann boundary condition $\eta=0$ are exactly the equations used to compute the expected resonant frequencies in Fig.~2. Equation \eqref{eq:ImodeCondition} corresponds to resonant modes in the red regions, \eqref{eq:OmodeCondition} to the blue regions, and \eqref{eq:AmodeCondition} for the yellow region. Note that unlike the yellow condition \eqref{eq:AmodeCondition}, the blue condition only takes the form \eqref{eq:OmodeCondition} in the case of severe evanescent separation. Additionally, one may encounter cases where $r_B<r_2$ for which there are no blue modes, and the red modes take their asymptotic form \eqref{eq:ImodeCondition}. If $r_B>r_2$ and the right side of \eqref{eq:IOmodeCondition} is not negligible (weak evanescent separation), then there is no natural concept of red and blue modes, but rather a set of combined resonant modes.

\end{appendices}

\end{document}